\documentclass{aa}  
\usepackage{graphicx}
\usepackage{rotating}
\usepackage{array}
\usepackage{subcaption}
\usepackage{caption}
\usepackage{soul}
\usepackage{mathtools}
\usepackage{txfonts}
\usepackage[dvipsnames]{xcolor}
\usepackage{natbib,twoopt}
\usepackage[breaklinks=true]{hyperref} 
\bibpunct{(}{)}{;}{a}{}{,}             
\makeatletter
  \newcommandtwoopt{\citepads}[3][][]{\href{http://adsabs.harvard.edu/abs/#3}%
    {\def\hyper@linkstart##1##2{}%
     \let\hyper@linkend\@empty\citep[#1][#2]{#3}}}
  \newcommandtwoopt{\citetads}[3][][]{\href{http://adsabs.harvard.edu/abs/#3}%
    {\def\hyper@linkstart##1##2{}%
     \let\hyper@linkend\@empty\citet[#1][#2]{#3}}}
  \newcommandtwoopt{\citeyearads}[3][][]%
    {\href{http://adsabs.harvard.edu/abs/#3}
    {\def\hyper@linkstart##1##2{}%
     \let\hyper@linkend\@empty\citeyear[#1][#2]{#3}}}
\makeatother
\begin{document}

   \title{Spectral characterisation of the extinction properties of NGC 3603 using JWST NIRSpec.}


   \author{Ciarán Rogers
          \inst{1},
          Bernhard Brandl
          \inst{1, 3}
          Guido de Marchi,
          \inst{2}}

   \institute{Leiden Observatory, Leiden University,
              PO Box 9513, 2300 RA Leiden, The Netherlands\\
              \email{rogers@strw.leidenuniv.nl}
         \and
             European Space Research and Technology Centre, Keplerlaan 1, 2200 AG Noordwijk, The Netherlands\\
             \email{gdemarchi@rssd.esa.int}
         \and
            Faculty of Aerospace Engineering, Delft University of Technology, Kluyverweg 1, 2629 HS Delft, The Netherlands}

   \date{Received November 10, 2023; accepted April 15, 2024}
   \titlerunning{The colour excess of NGC 3603 from JWST NIRSpec spectra.}
    \authorrunning{Ciarán Rogers, Bernhard Brandl, Guido de Marchi}

 
  \abstract
   {A necessary ingredient in understanding the star formation history of a young cluster is knowledge of the extinction towards the region. This has typically been done by making use of the colour-difference method with photometry, or similar methods utilising the colour-colour diagram. These approaches rely on adopting an extinction law with a given total-to-selective extinction ratio $R(V)$, or determining a value of $R(V)$ through empirical relationships. They also rely upon accurate spectral classification, reliable stellar isochrones, and separating field stars from genuine cluster members.}
   {The colour excess $E(B-V)$ can be independently determined by studying the decrements of the recombination lines produced by the nebular gas. Having access to many recombination lines from the same spectral series removes the need of adopting an extinction curve. Rather, different extinction curves can be trialled and the most appropriate one selected based on a minimum $\chi^2$ procedure.}
   {Using the Micro-Shutter Assembly (MSA) on board the Near InfraRed Spectrograph (NIRSpec), multi-object spectroscopy was performed, yielding 600 nebular spectra from the Galactic massive star formation region NGC 3603. The recombination line intensity ratios were used to determine independent values of $E(B-V)$. A series of extinction curves were trialled ranging from $R(V) = 2$ to $R(V) = 8$. The appropriate value of $R(V)$ was adopted based on the minimum $\chi^2$ procedure.}
   {The extinction characteristics of NGC 3603 are similar to other Galactic HII regions like Orion, as well as starburst regions such 30 Doradus in the Large Magellanic Cloud, in that we find a relatively large value of $R(V) = 4.8 \pm 1.06$, larger than the Galactic average of $3.1$. We find a typical value of $E(B-V) = 0.64 \pm 0.27$, significantly lower than values determined in previous studies. We also present a stacked nebular spectrum with a typical continuum S/N = $70$. This spectrum highlights the recombination lines of the HII region, several s-process elements such as Kr $III$ and Se $IV$, and molecular $H_2$ emission lines. This high S/N spectrum can act as a helpful template for identifying nebular emission lines.}
   {Using ratios of hydrogen recombination lines, we calculated the value of $R(V)$, $E(B-V)$ and $A(V)$ for $> 200$ lines of sight across NGC 3603. An extinction curve with a typical value $R(V) = 4.8 \pm 1.06$ is required to explain the colour excess observed in the nebular spectra. This corresponds to a typical $E(B-V) = 0.64 \pm 0.27$. This is significantly lower than what has been found in previous extinction studies of NGC 3603.}


   \maketitle
%
\section{Introduction}
Determining the age, distance, and star formation history of a stellar cluster relies upon an accurate knowledge of the extinction towards the region. Extinction is often characterised with the term $R(V) \equiv \frac{A_V}{E(B-V)}$, known as the total-to-selective extinction.
Most regions in the Milky Way follow a typical Galactic extinction law with a total-to-selective extinction ratio $R(V) = 3.1$. However, young regions with active ongoing star formation tend to exhibit higher values of $R(V)$, implying that the typical size of dust grains is large in these regions compared to the Galactic average. Indeed, many studies towards Galactic HII regions, or starburst galaxies, find systematically higher values of $R(V)$ \citep[e.g.][]{neckel1981interstellar, bautista1995nebular, de2014extinction, pandey2000stellar, calzetti2000dust} compared to the Galactic average of $3.1$. This could be a result of photoevaporation of smaller grains due to radiation from massive stars and/or the coagulation and growth from small to large grains, aided by the higher densities in star forming regions \citep{swamy1965grain}. \citet{gall2014rapid} show that supernovae explosions are also highly efficient means of producing large dust grains.\\ 
NGC 3603 is a giant Galactic HII region, $1-3$ million years old, located $7 \pm 1$ kpc away \citep{pandey2000stellar, melena2008massive, pang2016gray}. The region hosts an extremely compact OB cluster at its centre, which has made studying the central region in detail difficult due to limited angular resolution. With about one-third the mass in stars of 30 Doradus in the Large Magellanic Cloud (LMC), NGC 3603 is the closest system in the Galaxy to a genuine starburst region \citet{stolte2006secrets}. It is also the optically brightest HII region in the galaxy.\\
Most studies on the star formation history of NGC 3603 rely upon optical photometry, and make use of either the colour-difference method or two colour diagrams (TCDs) to determine the extinction \citep[e.g.][]{moffat1983bright, melnick1989galactic, pandey2000stellar, sung2004initial}, referred to as (Mof83), (Mel89), (Pan00), and (Sun04) going forward.\\
An alternative approach to determine the extinction law and colour excess towards a region is to study the recombination lines of nebular spectra. By comparing the ratios of hydrogen emission lines present in nebular spectra to their theoretical values based on Case B recombination theory \citet{osterbrock2006astrophysics}, the extinction law and colour excess can be determined, based on how much the measured ratios deviate from the theoretical ratios. In this study, we obtained $600$ nebular spectra using the multi-object spectroscopy (MOS) mode of NIRSpec on board the James Webb Space Telescope (JWST), using the G235H grating and F170LP filter. This corresponds to a resolving power of $\frac{\lambda}{\Delta \lambda} = 2700$, and a wavelength range of $1.6-3.2 \mu m$. MOS mode utilises the novel Micro-Shutter Assembly (MSA), which consists of $\sim 250\,000$ individual micro-shutters that can be configured open or closed depending on the positions of sources in the field of view.\\
In Sect. ~\ref{sec:sel_of_tar} we discuss the target selection. In Sect. ~\ref{sec:data_reduction}, the data reduction process is explained. In Sect. ~\ref{sec:ext_red}, the results of our extinction analysis are presented. We also present a high signal to noise ratio (S/N) stacked nebular spectrum in Sect. \ref{subsec:stacked_spectrum}. In Sect. \ref{sec:discussion}, we discuss our results, assumptions, uncertainties, and biases, and compare our results to the literature. In Sect. ~\ref{sec:conclusions} our findings are summarised.
\section{Selection of targets and observation strategy}
\label{sec:sel_of_tar}
\begin{figure*}[h]
    \centering
    \includegraphics[width=1\linewidth]{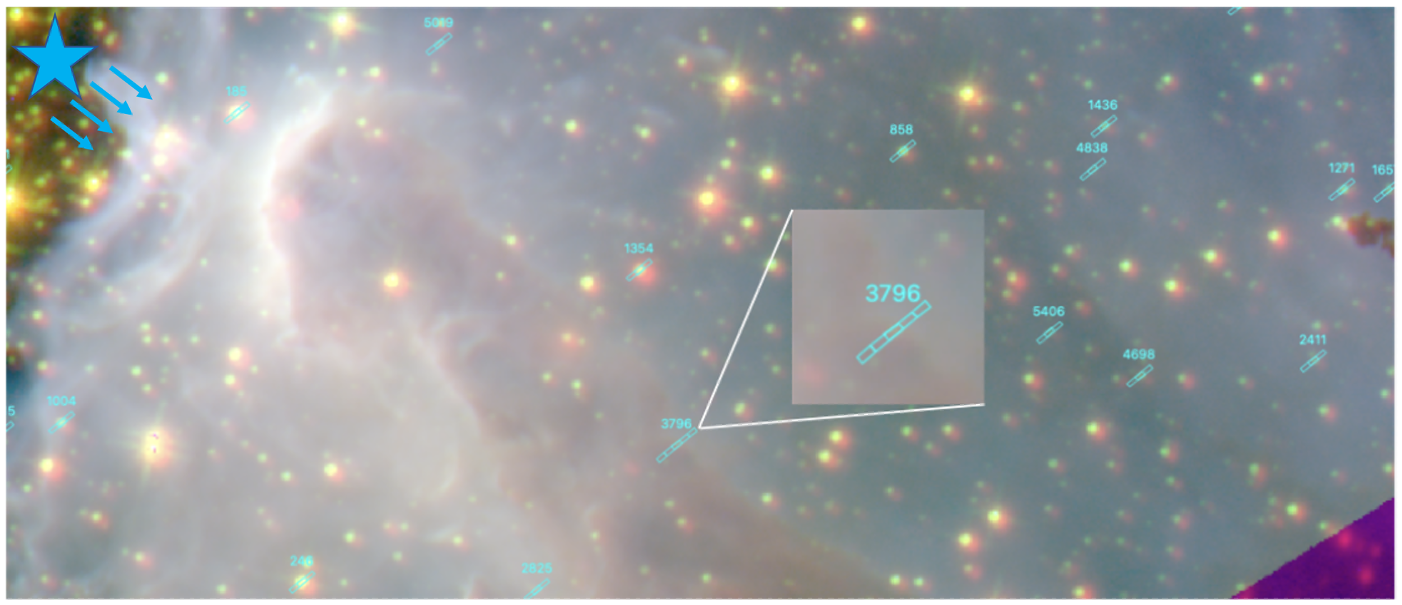}
    \caption{Region of NGC 3603 with the projected micro-shutter's angle and dimensions. This RGB image is composed of High Acuity Wide-field K-band Imager (HAWK-I) F$212n$ $H_2$ (Red), HST $F160W$ (Green), and HST $F168n$ (Blue). The stellar sources not discussed in this paper lie within the central micro-shutters, while the upper and lower shutters observe the nebula. The zoomed in section containing the source $3796$ illustrates the nodding pattern utilised. JWST nods up and down, moving three neighbouring open micro-shutters north-east and south-west, observing $6$ nebular regions in total. The position of the OB cluster is indicated by the blue star and arrows.}
    \label{fig:ngc_3603_zoom}
\end{figure*}

The principle aim of the NGC 3603 NIRSpec observations was to gain insight into the formation processes of the pre-main sequence (PMS) stars in the region. One hundred stellar sources were selected for observation with NIRSpec from an initial list of about 10000, originally observed with Hubble Space Telescope's (HST) Wide Field Camera 3 (WFC3), see \citet{beccari2010progressive}. The 100 sources observed with JWST had been classified photometrically as either PMS ($60/100$) or main sequence (MS) ($40/100$), based on $H{\alpha}$ excess emission. For our JWST observations, each source was placed within a micro-shutter, with the shutters directly above and below also open, forming a mini-slit of three micro-shutters. The upper and lower shutters were opened in order to measure the nebular background, which is crucial in a region like NGC 3603 where the nebular emission is bright. A three nod strategy was employed, moving the source from the central shutter, to the upper shutter, and finally to the lower shutter so that source spectra could be measured on different regions of the detector.\\
A major constraint on the final source selection was the placement of each source in a micro-shutter. Sources needed to be relatively isolated, so that contaminant stars were not observed in the neighbouring background shutters. Here a contaminant star is simply any star that happens to fall within one of the open micro-shutters, but is not a target of observation. This phenomenon was impossible to completely avoid in a region as dense with stars as NGC 3603. The impact of contaminants on our nebular spectra is discussed in the appendix, section \ref{subsec:contaminants}.\\ 
Figure \ref{fig:ngc_3603_zoom} shows a section of NGC 3603. The bright south-east pillar is visible, along with many projected micro-shutters. Note the sharp drop in nebulosity in the upper left hand corner of the image.
The final selection of 100 sources from the original catalogue of 10000 was primarily motivated by how isolated the sources were, so that contamination could be kept to a minimum. The final observation strategy consisted of three MSA configurations, with three nods per configuration. A full discussion of the MS and PMS stars will follow in Rogers et al. (in prep).\\ 
For this study, we focus on the nebular spectra that were obtained. Given the three nod strategy and two micro-shutters observing the nebular background per stellar source, six nebular spectra were obtained per source, totalling 600 nebular spectra in total.\\ 
Due to the nodding pattern of the telescope, certain nebular regions were observed multiple times, producing repeat spectra. The pointing accuracy of JWST is $~0.1\arcsec$ \citet{lallo2022jwst}. Given the distance to NGC 3603 of $7$ kpc, this small pointing offset results in sampling regions that are close in angular distance but relatively far apart ($\sim700$ AU) in terms of linear distance. For this reason we treat the repeat observations as independent spectra. The typical S/N in the continuum of each spectrum was $\sim10$. The recombination lines are quite strong, having a typical S/N of $\ge 30$, with the strongest lines $\ge 100$.
\section{Data reduction}
\label{sec:data_reduction}
\subsection{NIPS} \label{subsec:nips}
The uncalibrated data was largely reduced with the ESA Instrument Team’s pipeline known as NIRSpec Instrument Pipeline Software (NIPS) \citet{NIPS}. 
NIPS is a framework for spectral extraction of NIRSpec data from the count-rate maps, performing all major reduction steps from dark current and bias subtraction to flat fielding, wavelength and flux calibration and spectral extraction, with the final product being the 1D extracted spectrum. One of the final steps before extraction is the rectification of the spectrum. Since NIRSpec spectra do not run perfectly parallel with the dispersion direction, as well as bending slightly along the detector, a rectification step is performed on the so-called irregular 2D spectrum. This step resamples the 2D spectrum on a linear and uniform grid. The rectified spectrum, also called the regular 2D spectrum, is a count-rate map consisting of 3817 pixels in the dispersion direction and seven pixels in the cross-dispersion direction. The extraction step then simply collapses the 2D spectrum by summing the seven pixels along each detector column. During the data reduction process, it became apparent that contamination of the nebular spectra had occurred. This rendered the affected nebular spectra unusable. Contamination reduced the number of usable nebular spectra from $600$ to $471$. A full discussion of this effect is given in appendix \ref{subsec:contaminants}.\\

\section{Extinction and colour excess}
\label{sec:ext_red}
\subsection{Measured and theoretical line ratios}
It is possible to measure the extinction in a region via the recombination lines found in the nebular background spectra, \citep[e.g.][]{calzetti1997uv, calzetti1996dust}. Interstellar extinction reddens spectra, extinguishing blue wavelengths more than red wavelengths. The expected intensity ratio of two recombination lines for a given electron temperature $T_e$ and electron density $N_e$ is known from Case B recombination theory. The "B" signifies that the environment is optically thick to all Lyman series photons, that is, these photons are absorbed. The alternative scenario is Case A, where these photons are not absorbed \citet{osterbrock2006astrophysics}. Case B is expected to be valid for HII regions where the density of hydrogen is relatively high ($\ge 100 cm^{-3}$). The deviation that the measured recombination line ratios exhibit compared to Case B is a measure of the extinction that is taking place.\\
\subsection{The choice of recombination lines}
\label{choice_recomb_lines}
\begin{figure*}
    \centering
    \includegraphics[width=1\linewidth]{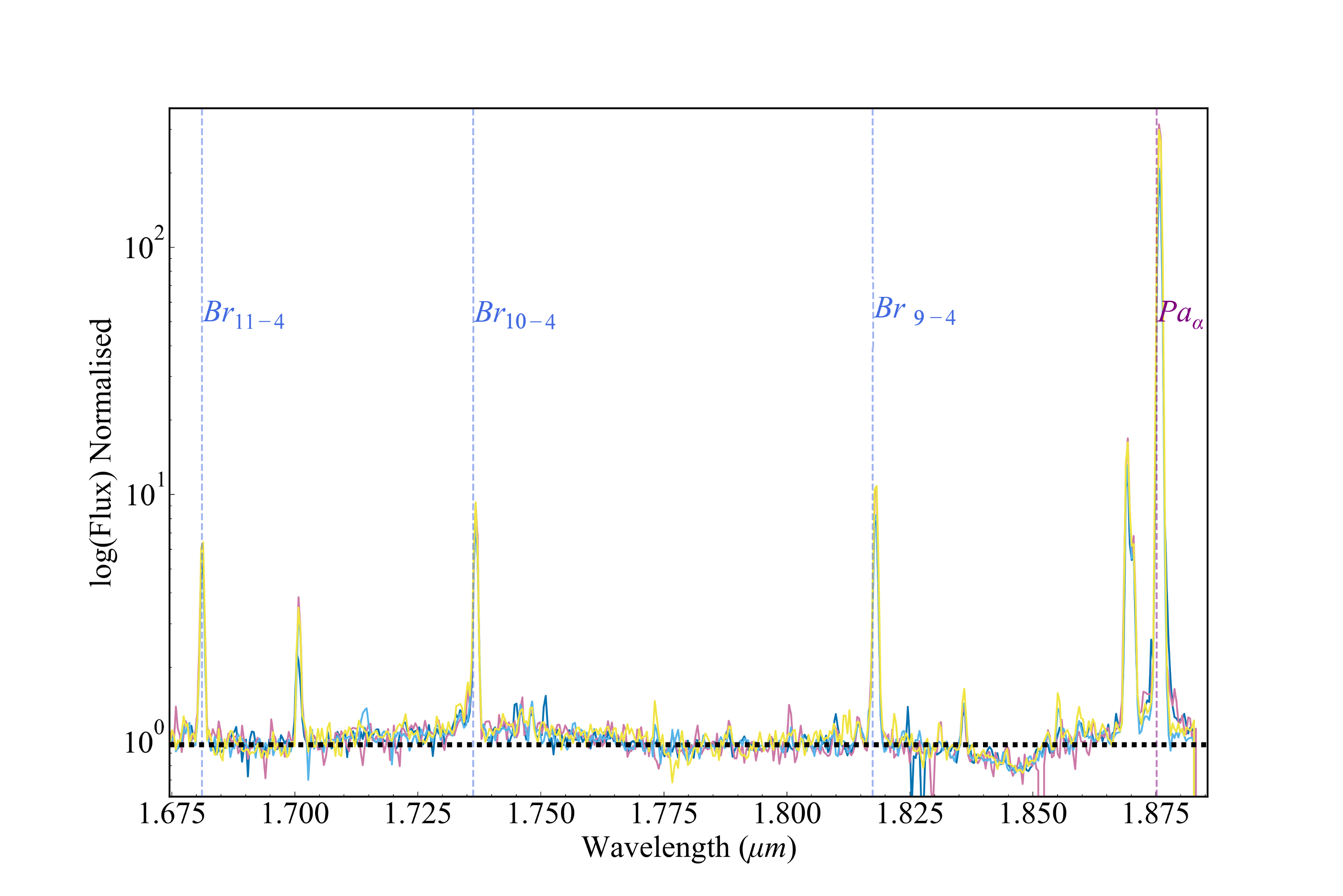}
    \caption{Example of four nebular spectra exhibiting a raised continuum level around $Br_{10-4}$ as a result of the NIRSpec calibration. This effect produces unphysical recombination ratios with respect to Case B recombination, and so this line has not been included in the remainder of the analysis.}
    \label{fig:calibration}
\end{figure*}

A strong Brackett series line in our nebular spectra could not be used in this extinction analysis - $Br_{10}$. This was because it is located in a region of the spectrum that is visibly affected by the calibration, and hence we could not measure a reliable line flux. Figure \ref{fig:calibration} illustrates this, showing four nebular spectra from neighbouring micro-shutters.
The continuum below the line is visibly raised, as indicated with the dashed black line tracing the correct continuum. The absolute change in the flux of $Br_{10}$ is a few $\%$. However, even a small change in the flux of the line can result in a significant change in the line ratio, which can significantly affect the resulting value of $E(B-V)$. As such, we have excluded $Br_{10}$ from this analysis.\\ 
All of the lines used in this analysis belong to the Brackett series. Using lines from the same series helps to ensure that the nebular conditions ($T_e$ and $N_e$) being probed are similar. Ideally all of the lines that we have used would have the same upper energy level, so that the excitation potential was the same, meaning that they arise from identical environments. This was not possible from the lines available to us, and so we opted to rely on a single hydrogen series, with all lines relatively close together in wavelength, keeping the physical conditions producing these lines similar. A recombination line list is given in table \ref{tab:line_list}.\\
\begin{table}
 \centering
 \begin{tabular}
{|p{1cm}|p{2cm}|p{2cm}|}
 \hline
 \multicolumn{3}{|c|}{Recombination lines} \\
 \hline
 Line & $\lambda$ ($\mu m$) & $\lambda^{-1}$ ($\mu m^{-1}$)\\
 \hline
 $Br_{11}$ & 1.6811 & 0.5948\\
 \hline
 $Br_{9}$ & 1.8274 & 0.5333\\
 \hline
 $Br_{8}$ & 1.9445 & 0.5143\\
 \hline
 $Br_{7}$ & 2.1661 & 0.4617\\
 \hline
 $Br_{6}$ & 2.6258 & 0.3808\\
 \hline
\end{tabular} 
\caption{\label{tab:line_list}Recombination lines used to estimate the extinction towards each source.}
\end{table}\\
\subsubsection{Using Pfund recombination lines to probe extinction}
There is a rich forest of Pfund recombination lines present in the nebular spectra. For a number of reasons, it has not been possible to study these lines to determine the extinction characteristics of NGC 3603. These lines lie at longer wavelengths than all of the Brackett lines except for $Br_{6}$. This means the extinction experienced by these lines is weaker, and hence harder to measure. The lines are also relatively close together in wavelength, making the differential extinction harder to measure. The lines are also weaker, resulting in larger uncertainties, which also hinders our determination of $E(B-V)$. As we shown in section \ref{subsubsec:det_ebv}, due to the overall weak differential extinction that we determine, the Pfund lines are not well suited to measuring $E(B-V)$ in NGC 3603, though we acknowledge that it would provide a very interesting secondary probe of extinction towards the region.
\subsubsection{Measuring the recombination lines}
\label{subsubsec:measure_recomb_lines}
\begin{figure*}[h]
    \centering
    \includegraphics[width=1\linewidth]{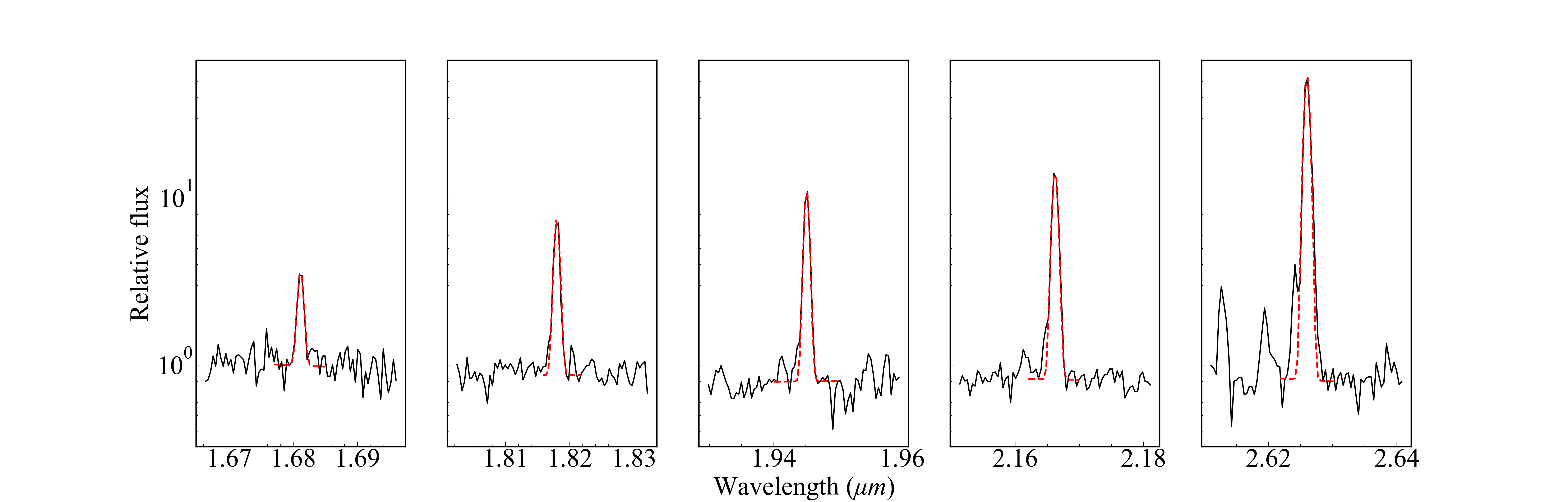}
    \caption{Five Brackett series line profiles are shown from a representative spectrum, plotted in log-scale. The best fitting Gaussian profile is shown as a red dashed line.}
    \label{fig:br_lines}
\end{figure*}

For each extracted nebular spectrum, we used Monte Carlo to estimate the line flux and uncertainty for the Brackett recombination lines, similar to that of \citep{ashraf2023h, riedel2017young}. This approach involved fitting the recombination lines many times, allowing the lines to vary within their uncertainties for each iteration. For this study we found that the recombination lines uncertainties were well sampled after $1000$ iterations. The flux of each line was determined by calculating the median of the $1000$ flux measurements, and the uncertainty of the flux was the standard deviation of the $1000$ measurements. We fit the recombination lines with Gaussian profiles using the non-linear least squares fitting routine "curve-fit" from SciPy. In combination with the analytical expression for a Gaussian, this fitting routine returns the best fitting values of the amplitude, standard deviation, and central wavelength of the Gaussian. We determined the equivalent width (EW) of the best fitting Gaussian profile, and converted the EW to a flux by multiplying by the level of the adjacent continuum. In some cases the hydrogen recombination lines were partially blended with another, weaker line. To estimate the impact of these weaker blended lines, we fit a double Gaussian to both lines and found that the relative flux contribution from the weaker lines was typically $<1\%$, so we do not consider it to significantly affect our results. An example of the line profiles of the Brackett series is shown in figure \ref{fig:br_lines}.\\
The flux of each hydrogen line was divided by the flux of $Br_{6}$, yielding a list of line ratios. The uncertainty of each ratio was found by propagating the error of each flux using the standard error propagation formula. The $Br_{6}$ line was chosen as the anchor (the line that we compare the other lines to) because it is present in all of the spectra (in some cases, $Br_{7}$ is located in the NIRSpec detector gap), and is always strongly detected, resulting in low statistical uncertainties. The measured line ratios then needed to be compared to the expected ratios from Case B recombination. To do this, the line ratios of \citet{storey1995recombination} were used for values of $T_e = 10000 - 12500 K$ and $n_e = 100-10000cm^{-3}$. The line ratios are not particularly sensitive to $T_e$ or $N_e$ and so our specific choice of these parameters is not extremely important. Our choice of $T_e$ comes from \citep{esteban2004carbon, mcleod2016connecting}, who both find typical electron temperatures of about $11000 K$ in NGC 3603.\\ 
Equation \ref{form1} gives the relationship between the colour excess, the line ratios, and the extinction curve,
\begin{gather} \label{form1}
    E(B-V) = \frac{-log(r_{obs}/r_{int})}{0.4[\kappa(\lambda_{line 1}) - \kappa(\lambda_{line 2})]} mag,
\end{gather}
where $r_{obs}$ is the measured ratio of two hydrogen recombination lines and $r_{int}$ is the theoretical ratio of two hydrogen recombination lines based on Case B recombination. We refer to the term: $log(r_{obs}/r_{int})$ as the ``decrement'' moving forward. The notation of formula \ref{form1} is taken directly from \citet{calzetti1996reddening}, and may lead to some confusion as the letter $r$ has been used to represent the line ratios, while we use it to represent the extinction curve in this study. In order to minimise confusion, we use lower case to differentiate between $r_{obs}$/$r_{int}$ - the line ratios, and $R(V)$ - the extinction curve. The terms $\kappa(\lambda_{line 1})$ and $\kappa(\lambda_{line 2})$ are equivalent to $R(\lambda_{line 1})$ and $R(\lambda_{line 2})$, that is, the values of the assumed extinction curve at the wavelengths of the recombination lines in question. We refer to this term as the ``extinction curve differences'' moving forward. Having numerous recombination lines meant that each set of ratios could provide an independent estimate of $E(B-V)$. The uncertainties of the line fluxes derived from the Monte Carlo iteration method as well as the uncertainties in $T_e$ and $N_e$ were propagated through to each value of $E(B-V)$ using the standard error propagation formulae.

\subsection{The choice of extinction curve}
If an inappropriate extinction curve is used in equation \ref{form1} to determine $E(B-V)$, different values of $E(B-V)$ will result from different line ratios. The appropriate extinction curve on the other hand, should produce the same value of $E(B-V)$ regardless of which line ratio is used, though this may not hold for lines with very large wavelength separations. This caveat is fully discussed in section \ref{sec:discussion}. 
The form of the NIR extinction curve that we have adopted is that of \citet{fitzpatrick2009analysis} (FM09),
\begin{equation} \label{form3}
    \centering
    \frac{E(\lambda - V)}{E(B-V)} = [0.349 + 2.087 R(V)] \times \frac{1}{1+(\lambda/0.507)^\alpha} - R(V),
\end{equation}
where $\alpha$ is the exponent that defines the curvature of the extinction curve, and $R(V)$ has the usual definition of $R(V) \equiv \frac{A_V}{E(B-V)}$. Previous studies of the Galactic extinction law in the NIR indicated that it could be represented as a universal power law with $A(\lambda) \propto \lambda^{-\beta}$, where the exponent $\beta$ is constant for any line of sight (the usually agreed upon value of $1.6 \le -\beta \le 1.8$), see \citep{draine1989interstellar, sneden1978infrared, schultz1975interstellar}. FM09 showed that $\beta$ is not only dependent on the line of sight, but that the wavelength range used to probe the extinction actually affects the value of $\beta$, and hence the value of $R(V)$ that is determined. See section $3.2$ of FM09 for a full discussion. Owing to its success in determining extinction towards a wide variety of regions, its flexibility, and simplicity, we have adopted the form of the extinction curve of FM09 using formula \ref{form3}. This extinction curve mimics a power law whose exponent increases with wavelength and can vary for different lines of sight, while keeping the number of free parameters to just two - $R(V)$ and $\alpha$. Here the exponent $\alpha$ is substituted for $\beta$ simply to highlight that it is no longer constant. By using the FM09 extinction curve, which has been characterised using only broadband photometry, we have assumed that it is appropriate for our wavelength range. This can be confirmed in the coming years with JWST spectroscopy.

\subsection{The $\alpha$-$R(V)$ relation}
\begin{figure}[h]
    \centering
    \includegraphics[width=1\linewidth]{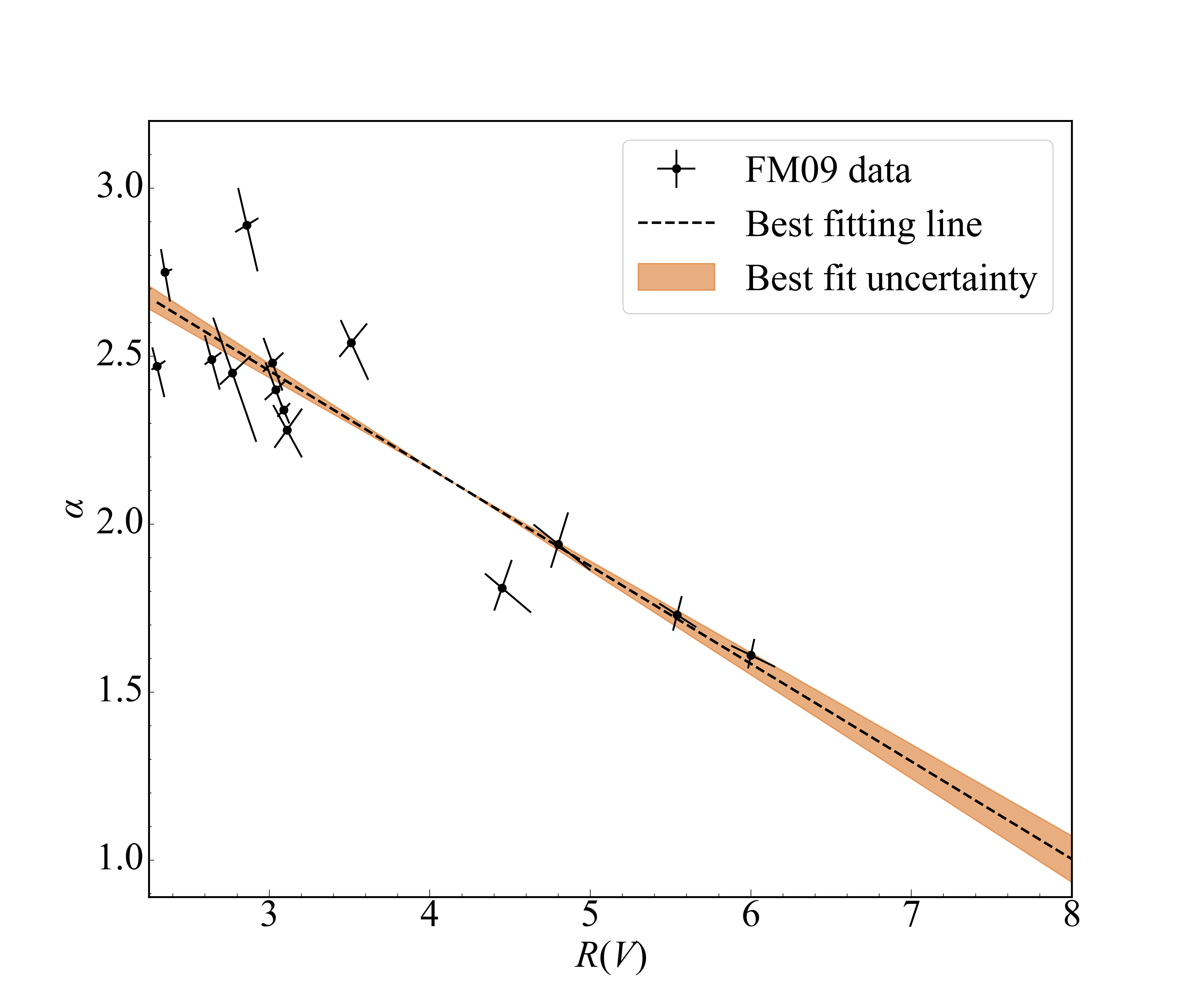}
    \caption{Relationship between $R(V)$ and $\alpha$ determined by \citet{fitzpatrick2009analysis}. The uncertainties in $\alpha$ and $R(V)$ are correlated, and the direction and strength of the error bars reflect that. The orange shaded band is the 1 $\sigma$ uncertainty of the fit, determined by performing 1000 Monte Carlo iterations.}
    \label{fig:alpha_Rv}
\end{figure}

Photometric extinction studies using high mass stars utilise both an expected colour excess $E(B-V)$ and expected brightness $M$ for each star. Both of these values are used in FM09 to determine $\alpha$ and $R(V)$. For nebular observations, while Case B recombination theory can provide $E(B-V)$ via equation \ref{form1}, there is no expected nebular brightness. This prevents us from directly determining $R(V)$ in equation \ref{form3}.
FM09 found that $\alpha$ and $R(V)$ are related, which allows us to determine $R(V)$ indirectly through $\alpha$. To do this, we took the values and $1 \sigma$ uncertainties for $\alpha$ and $R(V)$ from FM09, and ran Monte Carlo simulations to fit a line between the data points, and determine the uncertainty of the fitting parameters. The $\alpha$ - $R(V)$ relation is shown in figure \ref{fig:alpha_Rv}.
The uncertainties of $R(V)$ and $\alpha$ are correlated, as indicated by the direction of the error bars in figure \ref{fig:alpha_Rv}. We took care to only allow the data points to vary along the directions indicated by their error bars. 
We performed the Monte Carlo simulations multiple times,  fitting different types of curves to the data. These included a first- and second-order polynomial, as well as a power law. We found that these choices did not significantly affect the final result. The goodness of fits for each curve were very similar, and the best fitting curve closely resembled a straight line in all three cases. Hence, we chose to continue with the first-order polynomial. The equation of the line of best fit is

\begin{equation} \label{form_alpha_rv}
    \centering
    \alpha = -0.287 (\pm 0.018) \times R(V) + 3.309 (\pm 0.07).
\end{equation}
Using this relation, we replaced $\alpha$ in equation \ref{form1} with equation \ref{form_alpha_rv}, reducing it to a single parameter equation. Having done this, we were now able to begin fitting different extinction curves to solve for $R(V)$.

\subsection{Determining $R(V)$ and $E(B-V)$}
\label{subsec:find_alpha}
\begin{figure}[h!]
\begin{subfigure}{1\linewidth}
    \centering
    \includegraphics[width=1\textwidth]{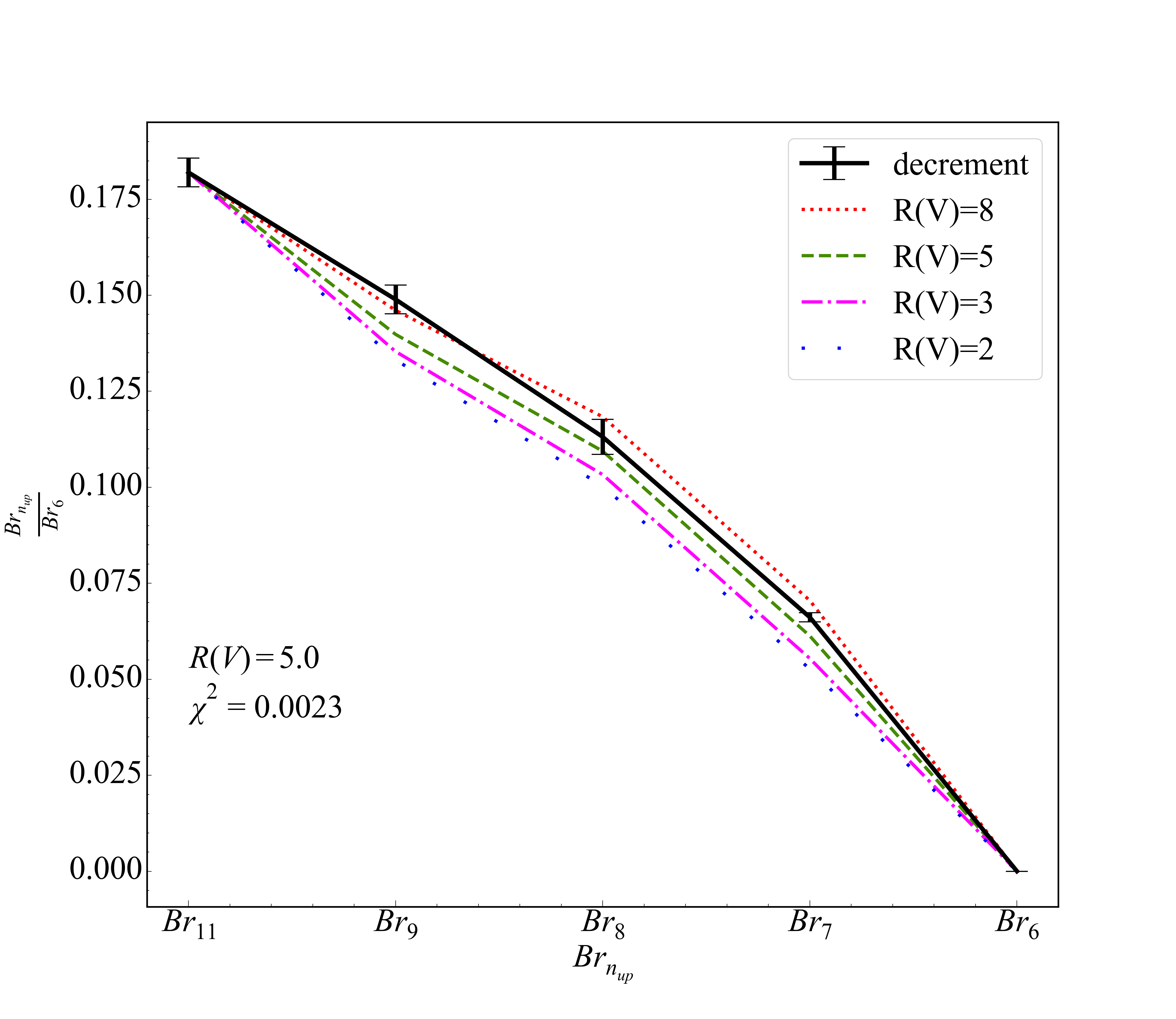}
    \caption{Set of decrements in which a reliable fit with low uncertainty can be found for the extinction curve with $R(V) = 5.0$.}
    \label{fig:dec_good}
\end{subfigure}
\begin{subfigure}{1\linewidth}
    \centering
    \includegraphics[width=1\textwidth]{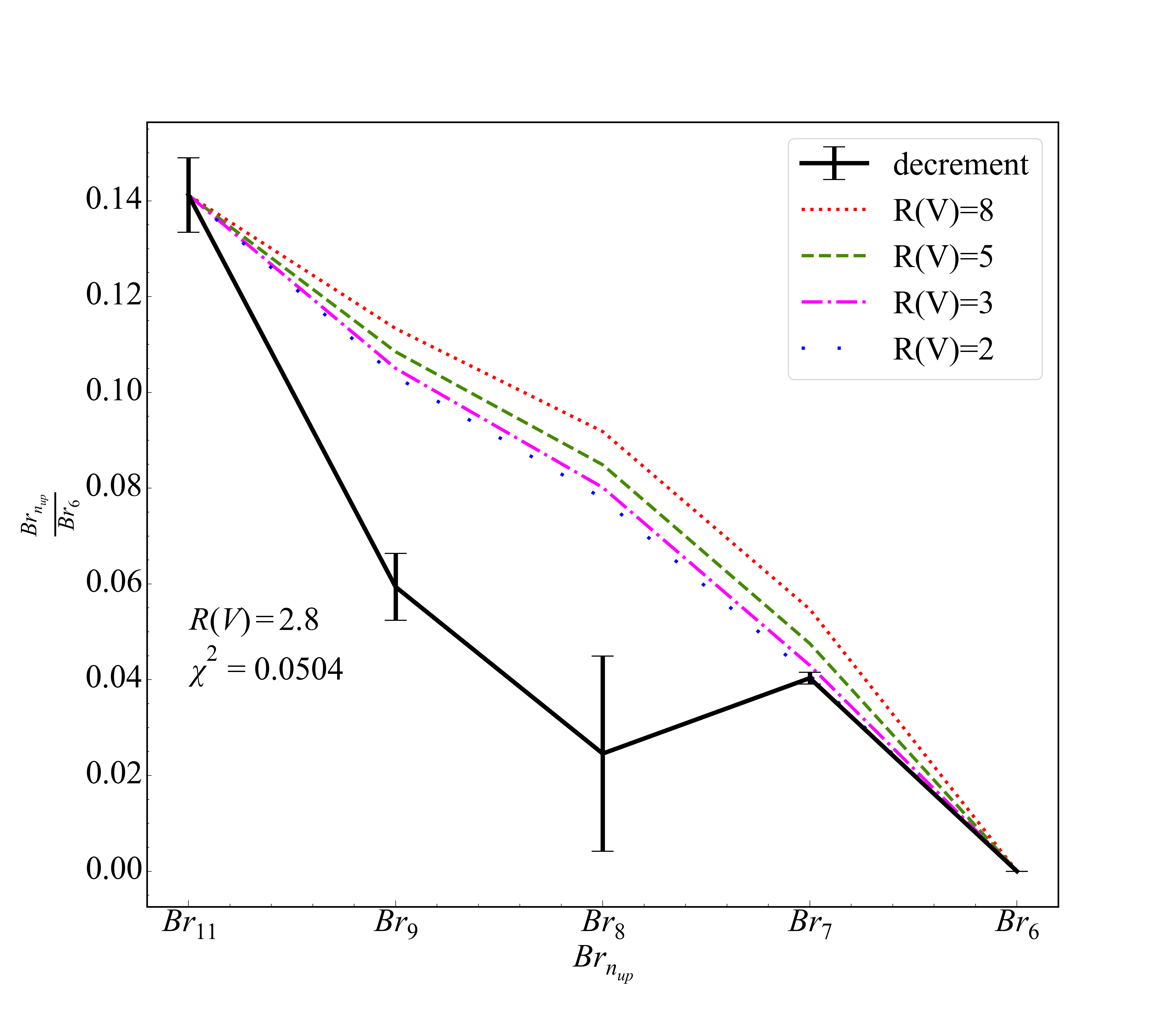}
    \caption{Set of decrements in which both the size of the error bars and the over all trend of the data points results in a poor fit for any extinction curves. The best fit is found for $R(V) = 2.8$}
    \label{fig:dec_bad}
\end{subfigure}
        
\caption{Two sets of decrements are shown as black solid lines. The uncertainties of each decrement are indicated with black error bars. Four normalised extinction curves are also shown. $R(V) = 2$ is represented by the blue, sparsely dotted line. $R(V) = 3$ is the magenta dot-dashed line. $R(V) = 5$ is the green dashed line, and $R(V) = 8$ is the red dotted line. Due to the relatively small differences in slope between extinction curves with $R(V) = 2$ and $R(V) = 8$ at these wavelengths, small uncertainties in the line ratios are required in order to reliably determine the best fitting value of $R(V)$}
\label{fig:dec_good_bad}
\end{figure}

Following equation \ref{form1}, in order to calculate $E(B-V)$, the decrement: $log(r_{obs}/r_{int})$ is divided by the extinction curve difference: $\kappa(\lambda_{line 1}) - \kappa(\lambda_{line 2})$. In order for each line ratio to produce the same value of $E(B-V)$, such that the dispersion is zero, the value of the decrements must be equal to, or be a constant multiple of the extinction curve differences. As such, by varying $R(V)$, the dispersion in $E(B-V)$ changes accordingly. By minimising the dispersion, the best fitting value of $R(V)$ can be found. The process of minimising the dispersion in $E(B-V)$ is analogous to performing the $\chi^2$ statistic test and minimising $\chi^2$, where $\chi^2 \equiv \Sigma\frac{(data_i-model_i)^2}{model_i}$.\\
Using equation \ref{form3}, we produced $100$ extinction curves with $R(V)$ going from $2$ to $8$. To determine the best fitting value of $R(V)$, each decrement was fit with all $100$ extinction curves, with each extinction curve being normalised to the level of the first decrement ($Br_{11}/Br_{6}$). By normalising, any offsets in the y-axis between the extinction curves are removed, and the difference between the decrements (data) and the extinction curve (model) are set entirely by the value of the exponent $\alpha$ in equation \ref{form3}, which we express in terms of $R(V)$ using equation \ref{form_alpha_rv}. Through this, by varying the value of $R(V)$, we could find the value that produced the minimum $\chi^2$ statistic.\\
To determine $R(V)$ for each decrement, we ran a Monte Carlo simulation, allowing the value of the decrements to vary within their uncertainties. In order to fully sample the uncertainties of the decrements, we needed to generate $10\,000$ realisations per spectrum. This larger number of iterations compared to section \ref{subsubsec:measure_recomb_lines} is due to the decrement uncertainties being larger than the line flux uncertainties. With each iteration, all $100$ normalised extinction curves were fit to the current realisation of the four decrements, calculating the $\chi^2$ statistic each time. The minimum $\chi^2$ and corresponding value of $R(V)$ were saved before moving on to the next realisation. This resulted in $471 \times 10\,000$ $\chi^2$s. We then calculated the median $\chi^2$ and $R(V)$ for each of the $471$ spectra. For all $471$ nebular spectra, a typical value of $R(V) = 4.9 \pm 1.5$ was found.\\ 
After careful inspection of the decrements with respect to their best fitting extinction curves, we chose to filter our sample of 471 lines of sight. This was based on a combination of low S/N spectra (emission lines with S/N$< 3$), resulting in large line ratio uncertainties, and spectra that produced unphysical jumps/spikes in the decrement trend. The reason for these jumps/spikes is still not clear to us, but may originate from unresolved issues with the NIRSpec calibration. We ordered the spectra in terms of increasing $\chi^2$, and chose to consider the best $200$. In general, beyond this point the decrements exhibited more unphysical spikes/jumps, with uncertainties becoming very large, of order $\ge 50\%$. An example of ``reliable'' and ``unreliable'' decrements are shown in figure \ref{fig:dec_good_bad}. From the $200$ most reliable decrements, we found the typical value of $R(V) = 4.8 \pm 1.06$. The nominal value of $R(V)$ is hardly altered by this filtering procedure, indicating that we have not introduced biases into $R(V)$ by only considering certain spectra. We also explored whether the smallest $\chi^2$ values corresponded to the brightest nebular spectra (with the highest S/N), as there is likely a correlation between brighter spectra and regions with lower extinction/colour excess. However, we did not find any significant correlation between the S/N of the spectra and their resulting $\chi^2$. As such, we felt it was justified to exclude the sight lines that increased our uncertainty in $R(V)$.\\ 
The $\chi^2$ statistic was used in this analysis as it is a familiar and readily understood measure of goodness-of-fit. However, the minimisation of dispersion in $E(B-V)$ has a more immediate physical meaning, in that the colour excess along the line of sight should be constant, and hence dispersion should be zero when the appropriate extinction curve is used. We performed the above analysis a second time, minimising the dispersion in $E(B-V)$ rather than $\chi^2$, and filtered our spectra based on increasing values of dispersion. By minimising the dispersion in $E(B-V)$, a typical value of $R(V) = 4.97 \pm 1.1$ was found, and $198/200$ spectra were overlapping between the two approaches, showing that the methods are effectively equivalent.

\subsubsection{Determining $E(B-V)$}
\label{subsubsec:det_ebv}
To determine the colour excess for each spectrum, we fit our $200$ spectra with their best fitting extinction curves. This was done via Monte Carlo as before, to propagate the uncertainties of the decrements and $R(V)$ to $E(B-V)$. This produced four values of $E(B-V)$ per spectrum. The median of these four values was calculated to determine the final colour excess for each spectrum. The median for all spectra was calculated, representing the typical value of $E(B-V)$ for our $200$ sight lines. We found a typical value of $E(B-V) = 0.64 \pm 0.27$.
\subsubsection{The total extinction $A(V)$}
Given the definition of $R(V) \equiv \frac{A(V)}{E(B-V)}$, having determined both $R(V)$ and $E(B-V)$, we were able to calculate the total extinction $A(V)$. We have found $A(V) = 3.1 \pm 1.5$ mag. This value represents both the foreground extinction due to the diffuse ISM, as well as the local extinction happening within NGC 3603. In order to separate these two distinct regimes of extinction, we have used the Gaia data release 3 catalogue \citet{brown2021gaia} of extinction values for $95$ sources located between $0.12^\circ - 0.5^\circ$ of NGC 3603, at a distance of $6-8$ kpc in order to determine the foreground extinction due to the diffuse ISM. We have obtained this data directly from ESA Sky. The visual extinction in the Gaia system, $A_0$, has a wavelength of $0.547 \mu m$, just slightly shorter than $A(V)$ at $0.551 \mu m$. Due to this small difference in wavelength we consider $A_0$ and $A(V)$ directly comparable. We have found a foreground extinction of $A_0 = 1.96$ mag. No measurement uncertainties for $A_0$ are provided in the Gaia catalogue. This gives a local extinction value of $A(V)_{local} = 1.1 \pm 1.5$ mag. This value is consistent within its uncertainties with zero. This suggests that all of the extinction that we have measured towards the region may be occurring in the foreground, and that the cluster itself does not contribute significantly towards the total extinction.

\subsection{Stacked nebular spectrum}
\label{subsec:stacked_spectrum}
A high S/N, stacked nebular spectrum has been created by combining and averaging all of the usable nebular spectra together, shown in figure \ref{fig:stacked_spec}. This process made use of all $471$ uncontaminated spectra. The spectra were first normalised, setting the continuum level to unity. They were then combined together by averaging each wavelength bin. In cases where the flux of a spectrum in a given wavelength bin deviated by $\ge 10\%$ of the median value, the wavelength bin for that spectrum was rejected. This removed noisy pixels, cosmic rays, and other spikes present in the spectra. The result was an average nebular spectrum for NGC 3603, with a typical continuum S/N = $70$. In this spectrum, the Paschen, Brackett and Pfund series are prominently in emission. The Pfund discontinuity can be seen clearly in this spectrum at $\lambda \ge 2.3 \mu m$. He I emission lines are also prominent throughout the spectrum, along with some weaker HeII emission. The ions Kr $III$ at $\lambda = 2.1986 \mu m$ and Se $IV$ at $\lambda = 2.287 \mu m$, produced by the s-process in Asymptotic Giant Branch stars are strongly detected in the nebular spectra. These ions are commonly seen in planetary nebula, \citep[e.g.][]{sterling2017identification}. There are also more than $40$ molecular hydrogen $H_2$ lines present, of which a selection have been annotated in figure \ref{fig:stacked_spec}. These lines are only seen in a subset of the nebular spectra. Typically these spectra correspond to regions of enhanced gas density, such as the south-east pillar shown in figure \ref{fig:ngc_3603_zoom}, though the lines are weakly detected in more diffuse regions as well.\\
This spectrum has been presented to showcase the rich NIR emission lines produced by the HII region, and we hope that it will be a useful template for identifying emission lines from other nebulae/HII regions.\\

\section{Discussion} \label{sec:discussion}
\subsection{Discussion of assumptions and uncertainties}
\label{subsec:assumps_uncerts}
In this study we have made several assumptions in order to determine $R(V)$ and $E(B-V)$ using formula \ref{form3}. Here a discussion of those uncertainties is given.
\subsubsection{Should $E(B-V)$ be constant for each line ratio?}
The basis of our analysis relies upon the correct extinction curve producing a constant value of $E(B-V)$, regardless of which line ratio is used to determine it. In \citet{bautista1995nebular} (B95), the authors measured the extinction $A(V)$ in Orion, using the nebular hydrogen emission lines from the Balmer, Paschen and Brackett series. They found that the value of $A(V)$ increased systematically with the mean wavelength of the series, with $A(V) = 1.03 \pm 0.05$, $A(V) = 1.22 \pm 0.10$, and $A(V) = 1.86 \pm 0.18$ for Balmer, Paschen and Brackett respectively. The authors suggest that this may be a result of optical effects due to the complex dust cloud geometry of the region. We also speculate that it may be a result of higher order spectral series (eg. Brackett, Pfund, Humphrey) coming preferentially from hotter/denser regions of the nebula, where the extinction properties may change. In (B95), the authors go on to say that the increase in $A(V)$ only becomes significant when the Balmer series is compared to the Brackett series (difference in $A(V) > 3\sigma$). For a single value of $R(V)$, the change in $A(V)$ must be accompanied by a proportional change in $E(B-V)$. Based on this, if spectral series separated by large wavelengths are used to probe the extinction properties of a region, then additional care must be taken to derive $R(V)$. We feel it is justified in our case to assume a constant $E(B-V)$ for each line of sight, as we make use of just a single spectral series.
\subsubsection{Calzetti equations assumptions}\label{subsubsec:calzetti_assumps}
We have used equation \ref{form3} to calculate $E(B-V)$, which itself relies upon two assumptions. These are: (1) the extinction that we are measuring is a purely foreground, uniform screen of dust, and (2) that our extinction law is appropriate. We know that (1) may not be completely valid in our case. The morphology of the gas and dust in NGC 3603 is visibly clumpy and irregular (though the intra-cluster extinction is consistent with zero from our results, albeit with large uncertainties). For a uniform foreground dust screen, all scattered and absorbed light is removed from the line of sight, with none of the photons being redirected back towards the telescope. The more complex geometry of dust clouds within NGC 3603 region means that scattered light can indeed be redirected back towards the telescope, including of course the emission lines that we use to measure the colour excess. We may therefore be measuring photons that have experienced scattering by dust, with blue photons being preferentially scattered compared to red photons, and obtain an underestimate of the amount of colour excess. The effect of scattering on extinction and colour excess due to different dust cloud geometries, degree of clumpiness, and aperture size has been investigated by \citet{krugel2009influence} (K09). One of their final conclusions is that scattering has a non-negligible impact on colour excess and extinction when the scattering optical depth $\tau_{sca}$ within the telescope aperture is larger than the relative observational error. The expression for $\tau_{sca}$ is given in (K09): $\tau_{sca} = \frac{1}{2} \phi D n \kappa_{sca}$ where $\phi$ is the spatial resolution of the telescope in radians, $D$ is the distance to the region in $cm$, $n$ is the number density in $cm^{-3}$ and $\kappa_{sca}$ is the scattering coefficient. Using this expression we can show that thanks to the exquisite resolution of JWST along with the relative proximity of NGC 3603 (compared to extragalactic sources), scattering should not significantly impact our results. For JWST at $\lambda = 2 \mu m$, $\phi = 0.064\arcsec = 3.15\times10^{-7}$ radians. The distance to NGC 3603 is $D = 2.16\times 10^{22} cm$, the number density is $10000 cm^{-3} \le n \le 100 cm^{-3}$ and the standard Milky Way value of $\kappa_{sca} \approx 2.5\times 10^{-22} cm^{-2}$. This gives ${8.4\times10^{-5} \le \tau_{sca} \le 8.4\times10^{-3}}$. The upper limit is approximately an order of magnitude smaller than our relative observational error of $\sim 6\times10^{-2}$, where our relative observational error refers to the typical relative uncertainty of each decrement. This indicates that the effect of scattering does not significantly affect our results. We acknowledge that without a complete radiative transfer computation, we cannot precisely quantify the effect of scattered light, but based on the estimate of $\tau_{sca}$ above, the impact should not be significant for JWST observations of NGC 3603.\\
For assumption (2), rather than adopting an extinction curve from the literature, we have tested a range of extinction curves, and so we are confident that this second assumption is satisfied in our analysis. A final point to note on the determination of $E(B-V)$, is that NGC 3603 has a very low Galactic latitude $\sim -0.5^{\circ}$ \cite{fukui2013molecular}, lying well within the Galactic plane. As such, it is possible that other Galactic star clusters may be incidentally located behind NGC 3603, which could affect the extinction properties that we determine. However, given that NGC 3603 is the optically brightest HII region in the galaxy, we feel it is a safe assumption that it is the dominant source of line emission along our lines of sight.
\subsection{Comparing $E(B-V)$ with the literature} \label{subsec:compare_ebv}
The value of $E(B-V) = 0.64 \pm 0.27$ that we have found is significantly lower than other estimates in the literature. In (Mof83), they found $E(B-V) = 1.44 \pm 0.09$ by spectrally classifying nine sources and using existing photometry to determine the colour excess. In (Mel89), they found $E(B-V) = 1.44 \pm 0.13$ within $50\arcsec$ of the central OB cluster. In that study they determine $E(B-V)$ with photometric data using the pair method, also known as the colour-difference method or MK method. This technique compares the observed colour and the theoretical colour of zero age main sequence stars that have been spectrally classified. They obtained a similar result using their nebular spectral data using the recombination lines, but critically, they assumed a Galactic extinction law with $R(V) = 3.2$. In (Pan00) they found $E(B-V) = 1.41 \pm 0.13$ for sources within $50\arcsec$ of the central cluster utilising the P-Q method from \citet{ogura1981ubv}. They derive a value of $R(V) = 4.3 \pm 0.2$ for NGC 3603.
When they included NIR photometry from \citet{eisenhauer1998stellar} the value of $R(V)$ increased to $4.2 \pm 0.7$, $4.8 \pm 0.8$, and $5.2 \pm 1.0$, for the J, H and K filters respectively. The uncertainty was much higher for these NIR observations, and so (Pan00) adopt their own value of $R(V) = 4.3 \pm 0.2$ owing to its lower uncertainty. 
In (Sun04), it was found that $E(B-V) = 1.25$ using the colour-difference method, and using the relation \ref{form4} from \citet{guetter1989reddening},
\begin{equation} \label{form4}
    \centering
    R(V) = 2.45 \times \frac{E(V-I)}{E(B-V)},
\end{equation}
they found a value of $R(V) = 3.55 \pm 0.12$. They also determined a distance modulus $V_0 - M_V = 14.2 \pm 0.2$ for NGC 3603 based on their estimate of the colour excess.
\citet{pang2016gray} (Pang16) also found a value of $R(V)$ that is higher than the typical galactic value. Within the filter system for HST, they found $R_{f555w} = 3.75 \pm 0.87$ and a value of $E(f435W - f555W) = 1.33 \pm 0.12$ mag. They do this using the pair method to determine $E(f435W - f555W)$, and perform a similar step to determine the total extinction $A_{f555W}$ by comparing the measured magnitude and theoretical magnitude of zero age main sequence stars. This allows them to directly calculate $R_{f555W}$ from the relation $R_{f555W} \equiv \frac{A_{f555W}}{E(f435W - f555W)}$. We note that they rely upon the distance modulus $V_0 - M_V$ from (Sun04) to determine the total extinction, but the distance modulus from (Sun04) was determined based on a different colour excess law than that found by (Pang16). See \citet{melena2008massive} for a full summary of selective extinction values from literature for NGC 3603.

\subsection{Spatial sampling of nebular spectra}
A possible explanation behind why our value of $E(B-V)$ is so much lower than what has been found in previous studies may be the spatial distribution of the nebular spectra in our sample. The extinction studies mentioned in \ref{subsec:compare_ebv} make use of the colours of massive stars that are close to the cluster centre ($\le 50\arcsec$). Our nebular spectra sample a wider area, with the majority of our spectra coming from regions $\ge 60\arcsec$ from the cluster core, and some spectra coming from as far as $\sim100\arcsec$ from the cluster core. Many of our lines of sight sample denser regions of the nebula, where large clouds of gas and dust can be seen. This is qualitatively different from the inner $50\arcsec$ of the cluster, where the density of gas and dust is much lower. This is evident in figure \ref{fig:ngc_3603_zoom}, in the top left of the image, where the nebular emission drops off sharply beyond the south-east pillar. This variation in gas and dust density could lead to different extinction properties across the region. Indeed, we have found that the best fitting value of $R(V)$ and resulting $E(B-V)$ varies across the region, with a $1 \sigma$ spread of $\pm 1.25$ and $\pm 0.6$ respectively. As such, we believe it is likely that our lower typical value of $E(B-V)$ is not contradicting previous studies, rather it is reflective of the different regions of the nebula that we are probing.

\section{Conclusions}
\label{sec:conclusions}
Using the NIR nebular spectra of NGC 3603, we have determined a new extinction law towards the giant HII region, mostly corresponding to regions $\ge 50\arcsec$ from the cluster centre. We have done this by exploiting the strong Brackett recombination lines present in our nebular spectra obtained with JWST NIRSpec. By fitting a range of extinction curves to each spectrum, we were able to home in on the most appropriate extinction curve, which has the form of FM09's NIR extinction law. We have found a typical value of $R(V) = 4.8 \pm 1.06$ for the region as a whole. This corresponds to a colour excess of $E(B-V) = 0.64 \pm 0.27$, which is about half the value published in previous works. Our findings diverge from what has been found previously, but we believe this is a result of the spatial sampling of the region, and highlights how spatially variable the extinction law is across the HII region.\\
We have also created a stacked nebular spectrum by averaging together all of the usable individual nebular spectra, resulting in a typical continuum S/N = $70$. This stacked spectrum highlights the emission lines produced by the nebula of NGC 3603, and can be used as a template for identifying otherwise hard to detect emission lines in fainter nebulae.
\section*{Acknowledgements}
We thank our referee for their suggestions and comments which have greatly improved both the analysis presented here, as well as the readability and flow of the paper.
\bibliographystyle{aa} 
\bibliography{references.bib} 

\appendix
\section{Impact of contaminant stars} \label{subsec:contaminants}
\begin{figure*}[h]
    \centering
    \includegraphics[width=1\linewidth]{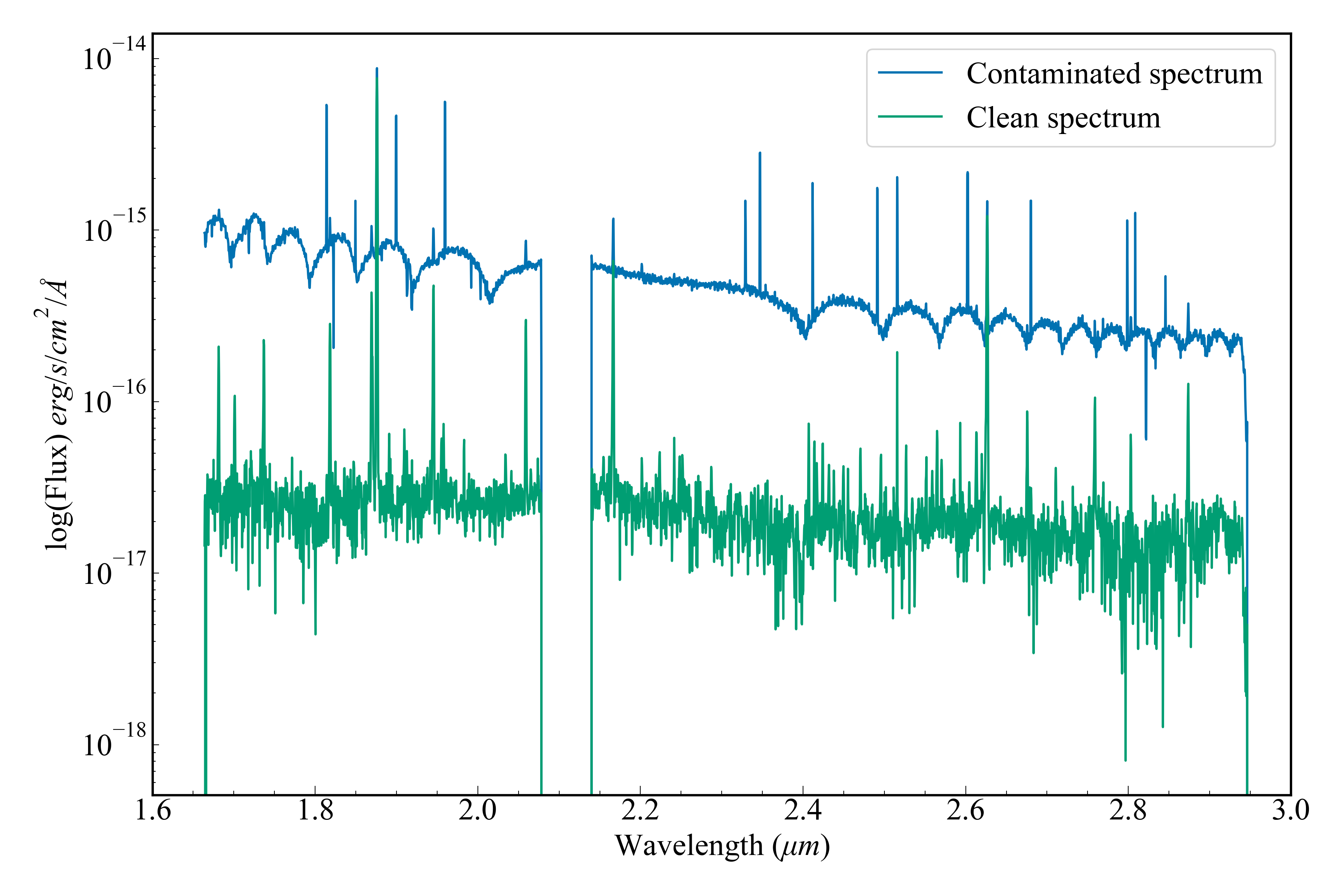}
    \caption{Two nebular spectra, one that has been extracted from a clean nebular region free of contaminants, and another that is dominated by a contaminant.}
    \label{fig:contaminant_spectrum}
\end{figure*}

Of the $600$ nebular spectra obtained with NIRSpec, $129$ were not usable due to contamination by an incident star, leaving us with $471$ nebular spectra. As mentioned in section \ref{sec:sel_of_tar}, a limitation on the number of sources that could be observed in a single MSA configuration came from the intense crowding of NGC 3603. In order to ensure that only well isolated sources were observed, with clean neighbouring background shutters, a criterion was implemented which specified that a source could be considered for observation so long as any other nearby stars were at least five magnitudes less bright in the V band. This approach was successful in so far that every target star had at least one clean background shutter. However, the presence of contaminant stars, even faint ones, has a significant impact on the extracted nebular background spectrum. Figure \ref{fig:contaminant_spectrum} shows an extracted spectrum from a nebular background shutter that contained a contaminant star, along with a spectrum from a clean neighbouring shutter.
In figure \ref{fig:contaminant_spectrum} the upper blue spectrum is an example of a contaminated nebular spectrum. The lower green spectrum is an example of a clean nebular spectrum, with no contamination. Both of these spectra come from neighbouring micro-shutters, and so they should look quite similar to each other in terms of over all brightness and spectral slope. The most obvious difference between the clean and the contaminated spectra is the shape of the continuum, which for the contaminated spectrum, takes on a quasi-periodic oscillating profile. It is not clear to us why this happens, however it is apparent that this occurs only in cases where there is a contaminant star. This phenomenon was also seen in previous work with simulated NIRSpec data \citet{rogers2022quantifying}. With the simulated data, contaminants could be added or removed at will, and the same oscillating pattern disappeared from the nebular spectra once the contaminants were removed. It is also clear in the 2D spectrum of affected nebular spectra, where a narrow, bright spectral trace can be seen overlaid on top of the extended nebular emission.\\ 
We have attempted extracting contaminated spectra in different ways to try to minimise or remove the effect of the contaminant. NIPS allows us to extract the spectrum in multiple ways. The traditional approach is to simply sum the flux in each detector column, but it is also possible to take a median value of the flux in each column, and multiply this value by the number of pixels per column (in our case for the 2D rectified spectra, this is seven pixels). The benefit of using the median approach for nebular spectra is that, under the assumption that the nebula is spatially uniform across the micro-shutter, the S/N of the extracted spectrum can be improved. This is because there is some slit loss around the edges of the shutter which can essentially be corrected by taking the median flux value. In the case where the contaminant spectrum is located near the upper or lower edge of the spectral trace, the median pixel value would correspond to the nebular light, excluding the contaminant flux. In the more common case of bright contaminant stars, whose spectral trace can cover $\ge 50\%$ of the seven pixels per column, the stellar flux dominates the spectrum, and taking the median effectively considers only the stellar flux. For this reason we opted to stick with the traditional sum method.\\ 
Another approach to remove the contaminant flux is to simply not count the detector rows that contain the stellar flux. Rectifying the spectrum means that the spectral trace runs perfectly parallel with the dispersion direction, making it relatively simple to identify which rows contain the contaminant flux, and skip them. Unfortunately, the resampling of the flux during rectification does not work perfectly, with some pixels receiving more flux than they should, and others receiving too little. The result is that extracting only a portion of the nebular spectrum - the portion that does not contain contaminant flux, results in its own oscillating continuum due to the resampling. This is most easily seen in data like ours, with high S/N and high resolution. Based on this, we opted to simply exclude any nebular spectrum that contained a contaminant. This brought the number of usable nebular spectra from $600$ to $471$.\\

\newpage
\section{Stacked nebular spectrum}
\begin{sidewaysfigure*}[h]
    \centering
    \includegraphics[width=\textwidth,height=\textheight,keepaspectratio]{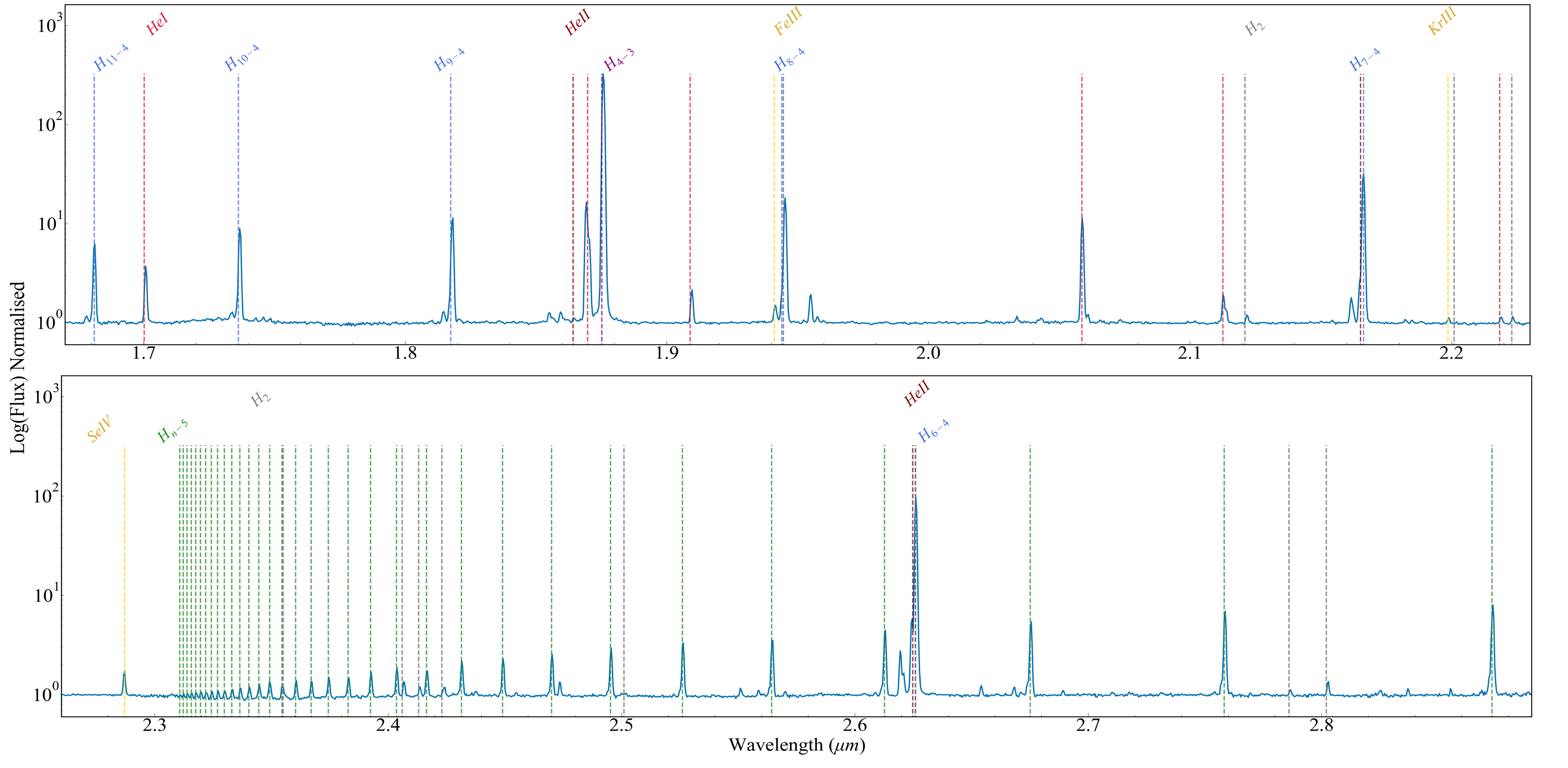}
    \caption{Stacked spectrum, resulting from the averaging and combining of the individual nebular spectra.}
    \label{fig:stacked_spec}
\end{sidewaysfigure*}

\end{document}